\documentclass[prl,a4paper,twocolumn,showpacs,superscriptaddress,floatfix]{revtex4}
\usepackage{times}

\usepackage{graphicx}
\usepackage{amsmath}
\usepackage{amssymb}
\usepackage{bm}
\newcommand{\be}{\begin{equation}}
\newcommand{\ee}{\end{equation}}

\begin{document}

\title{Time-Reversal Symmetry Breaking and Decoherence in Chaotic Dirac Billiards}

\author{A. J. Nascimento J\'unior$^1$, M. S. M. Barros$^2$, J. G. G. S. Ramos$^2$, A. L. R. Barbosa$^1$}

\affiliation{$^1$ Departamento de F\'isica, Universidade Federal Rural de Pernambuco, 52171-900 Recife-PE, Brazil\\ $^2$ Departamento de F\'sica, Universidade Federal da Para\'iba, 58051-970 Jo\~ao Pessoa Para\'iba, Brazil}

\date{\today}

\begin{abstract}

In this work, we perform a statistical study on Dirac Billiards in the extreme quantum limit (a single open channel on the leads). Our numerical analysis uses a large ensemble of random matrices and demonstrates the preponderant role of dephasing mechanisms in such chaotic billiards. Physical implementations of these billiards range from quantum dots of graphene to topological insulators structures. We show, in particular, that the role of finite crossover fields between the universal symmetries quickly leaves the conductance to the asymptotic limit of unitary ensembles. Furthermore, we show that the dephasing mechanisms strikingly lead Dirac billiards from the extreme quantum regime to the semiclassical Gaussian regime.

\end{abstract}


\maketitle

\section{Introduction}
\label{}

\begin{figure*}[!]
\centering
\includegraphics[width=0.49\textwidth]{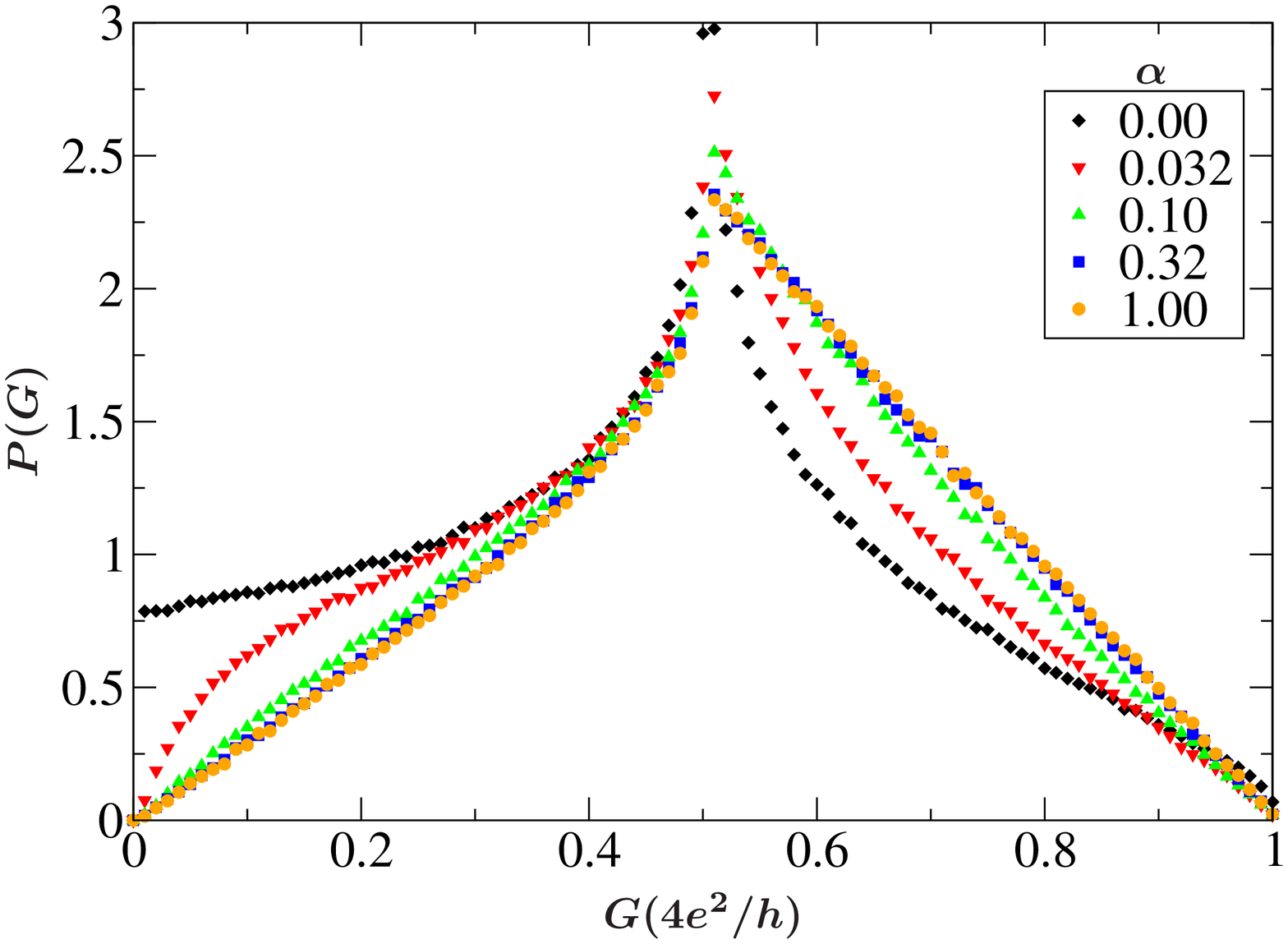}
\includegraphics[width=0.49\textwidth]{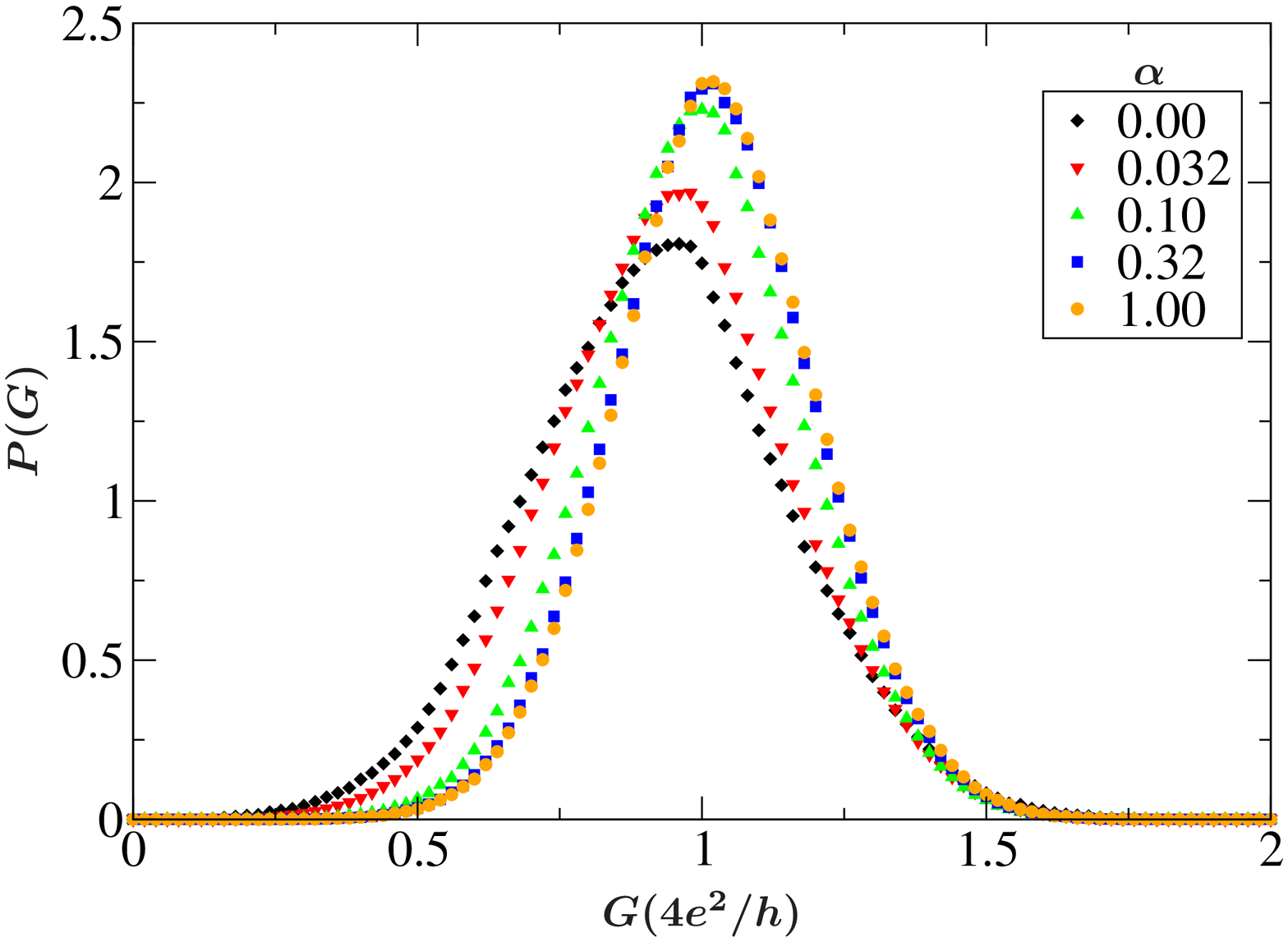}
\caption{Probability distribution of conductance for chaotic Dirac billiard in the extreme quantum regime (left), $N_1=N_2=1$, and the intermediate regime (right), $N_1=N_2=2$, in the presence of a phase coherence breaking parameter $N_\phi=0$. For $\alpha =0,1$ the distributions correspond to pure ensembles chGOE and chGUE, respectively.}\label{figura1}
\end{figure*}

Extensively tested, the random matrix theory (RMT) has enjoyed success in describing the electronic transport properties on disordered mesoscopic systems \cite{Berry,Bohigas,Stockmann,ARichter,MelloKumo,beenakker,Dietz,BeenakkerMarcus}. Following the classification of Cartan, the RMT is divided into ten symmetry classes: three Wigner-Dyson, three Chiral and four Altland-Zirnbauer ensembles \cite{Zirnbauer,JacquodButtiker}. The Wigner-Dyson ensembles \cite{mehta} are used to describe the electronic transport in mesoscopic devices with or without time-reversal (TRS) and spin-rotation  symmetries (SRS). We call their chaotic quantum dots counterparts as chaotic Schr\"odinger billiards. Furthermore, the Chiral ensembles \cite{Verbaarschot} are used in systems with sublattice/chiral symmetry, with or without TRS and SRS, and we will refer to them as chaotic Dirac billiards  \cite{Mondragon,Geim,Mortessagne,Baranger,Grafeno,Barros,Klaus}. The Altland-Zirnbauer ensembles are used in quantum electronic devices connected to superconductors \cite{Zirnbauer}. Henceforth, this last ensemble will be called, by analogy with the previous one, chaotic Andreev billiards \cite{beenakkerAndreev}.

The RMT applied to chaotic Schr\"odinger billiards has been widely discussed in the literature \cite{MelloKumo,beenakker,Dietz}. The theoretical works covering this topic include universal fingerprints as the probability distributions  of transmission eigenvalues and conductance \cite{Mello1,Lenz, Brouwer1,Kanzieper,Souza}, interference effects \cite{Ramos,nos,Dietzprb,paradox,nos1,Novaes}, entanglement \cite{BeenakkerMarcus,Kanzieper,Gopar,Almeida}, time-reversal symmetry breaking \cite{Lenz,Souza,Berry2,Schanze,Dietz3} and phase coherence breaking or decoherence \cite{Buttiker,Marcus,Mello,Brouwer,Petitjean}, which produce a large number of experimental fundamental consequences \cite{MelloKumo,beenakker,Dietz,BeenakkerMarcus}. However, the analytical results are usually obtained in very specific regimes \cite{MelloKumo,beenakker}, motivating numerical studies of regimes beyond such analytical limitations and giving a deep understanding of electronic transport properties \cite{ Brouwer1,Souza,Almeida,Schanze,Lewenkopf}.

Recently, some studies have been focusing on chaotic Dirac devices \cite{Mondragon,Geim,Mortessagne,Baranger,gafeno,Barros,Klaus,Wehling,Silva, Baranger, Grafeno, Baranger1, Adagideli, Richter, Richter1,Stampfer,Grebogi,Rycerz,Dietz4,Dietz5,LeiYing} due to the modern control of novel materials such as graphene and topological insulators structures. Using the tight-binding Hamiltonian model, Ref.\cite{Markos} analyzes numerically the density of states and probability  distributions of conductance for a disordered two-dimensional electron systems with chiral symmetry at zero energy. More recently, Ref.\cite{Barros} uses the diagrammatic method to obtain analytical expressions for both the mean and the variance of chaotic Dirac billiards conductance. This diagrammatic method also analyzes the mechanisms of phase coherence breaking in the semiclassical regime (large number of open channels). However, it is not a simple task to obtain analytical results for the crossover regime when there is a single open channel (defined as extreme quantum regime) in the leads coupled to the quantum dot. Therefore, it is necessary to perform numerical analysis on the crossover regime. The analytical difficulty occurs precisely due to the peculiarities of the probability distribution of conductance in the crossover regime, which has an extreme deviation from the Gaussian \cite{Mello1,Souza} and, consequently, forbids the integrability.

The aim of this paper is to present a complete numerical analysis of the time-reversal symmetric breaking and the phase coherence breaking mechanisms in the extreme quantum regime of chaotic Dirac billiards. Furthermore, we analyze the competition between both mechanisms over three remarkable aspects of conductance: the probability distribution, the weak localization and the universal fluctuations. With this motivation, we adapt to the Chiral ensembles the method established in the Ref.\cite{Souza}, which was developed to study the crossover regime (time-reversal symmetric breaking mechanism) between the Gaussian orthogonal ensemble (GOE) and the Gaussian unitary ensemble (GUE) of the Wigner-Dyson universal classes.

\section{Scattering Model and Decoherence}
\label{}

The study of quantum dots (QD) is usually performed with the use of two leads, one connected to a source and the other one to an electron drain. We consider the more general configuration of a QD connected to three leads, the third one being used to theoretically introduce the mechanisms of decoherence. In this section, we present the standard Hamiltonian scattering model for the general configuration. Firstly, the Sub-lattice/Chiral Symmetry is incorporated by the massless Dirac Hamiltonian, which satisfies the following anti-commutation relation \cite{Verbaarschot}
\begin{eqnarray}
\mathcal{H}= -\sigma_{z}\mathcal{H}\sigma_{z}, \quad
\sigma_{z}=
\left[
\begin{array}{cc}
\textbf{1}_{M} & 0\\
0 & -\textbf{1}_{M}
\end{array}
\right].\label{H}
\end{eqnarray}
The $\mathcal{H}$-matrix has order $2 M$ and $\textbf{1}_{M}$ is an $M \times M$ identity matrix. From a physical point of view, we can interpret the $M$ number of $1$'s and $-1$'s in $\sigma_z$ as the number of atoms in each sublattice \cite{Barros,Silva}, in a total of $2M$ atoms in the chaotic Dirac billiard. The  Hamiltonian model for the scattering matrix, $\mathcal{S}$, can be written as \cite{ARichter}
\begin{eqnarray}
\mathcal{S}(\epsilon)=\textbf{1}-2\pi i\mathcal{W}^{\dagger}(\epsilon-\mathcal{H} +i\pi\mathcal{W}\mathcal{W}^{\dagger})^{-1}\mathcal{W}.\label{MW}
\end{eqnarray}
The $\mathcal{S}$-matrix has order $\bar{N}_T$, where  $\bar{N}_T=\bar{N}_1+\bar{N}_2+\bar{N}_3$ is the total number of open channels or atoms in the three leads connected to the chaotic Dirac billiard. In each lead, there are $\bar{N}_i =2N_i$ open channels ($i=1,\dots,3$). In other words, there are $N_i$ open channels in each sub-lattice and, as the system is compounded by two sub-lattices, there are $2N_i$ open channels in each lead. The $2M \times \bar{N}_T$ matrix $\mathcal{W}$ represents all coupling combinations of the chaotic Dirac billiard resonances to the open channels of the leads. The scattering matrix is unitary $\mathcal{S}^\dagger\mathcal{S}=\textbf{1}$ due to the conservation of the electronic charge. Equations (\ref{H}) and (\ref{MW}) demonstrate that the $\mathcal{S}$-matrix satisfies the relation
\begin{eqnarray}
\mathcal{S}=\Sigma_{z}\mathcal{S}^{\dagger} \Sigma_{z}, \quad
\Sigma_{z}=
 \left[\begin{array}{cc}
\textbf{1}_{N_T}  & 0\\
0 & -\textbf{1}_{N_T}
 \end{array}
 \right] \label{S}
\end{eqnarray}
at the Dirac point, i.e., at zero energy ($\epsilon = 0 $). The $\mathcal{S}$-matrix is conveniently written as a function of transmission, $t$, and reflection, $r$, blocks as
\begin{eqnarray}
\mathcal{S}=
 \left[\begin{array}{ccc}
r_{11}  & t_{12} & t_{13}\\
t_{21} & r_{22} & t_{23}\\
t_{31} & t_{32} & r_{33}
 \end{array}
 \right]
\end{eqnarray}
where $t_{ij}$ and $r_{ij}$ have dimension $\bar{N}_i \times \bar{N}_j$, with $i,j=1,\dots,3$. In fact the Hamiltonian describing the Dirac Billiard is evaluated for energies (Dirac) close to the Fermi energy. Having in mind the homogeneity of the spectrum, we have fixed the Fermi energy to be zero.

\begin{figure*}[!]
\centering
\includegraphics[width=0.49\textwidth]{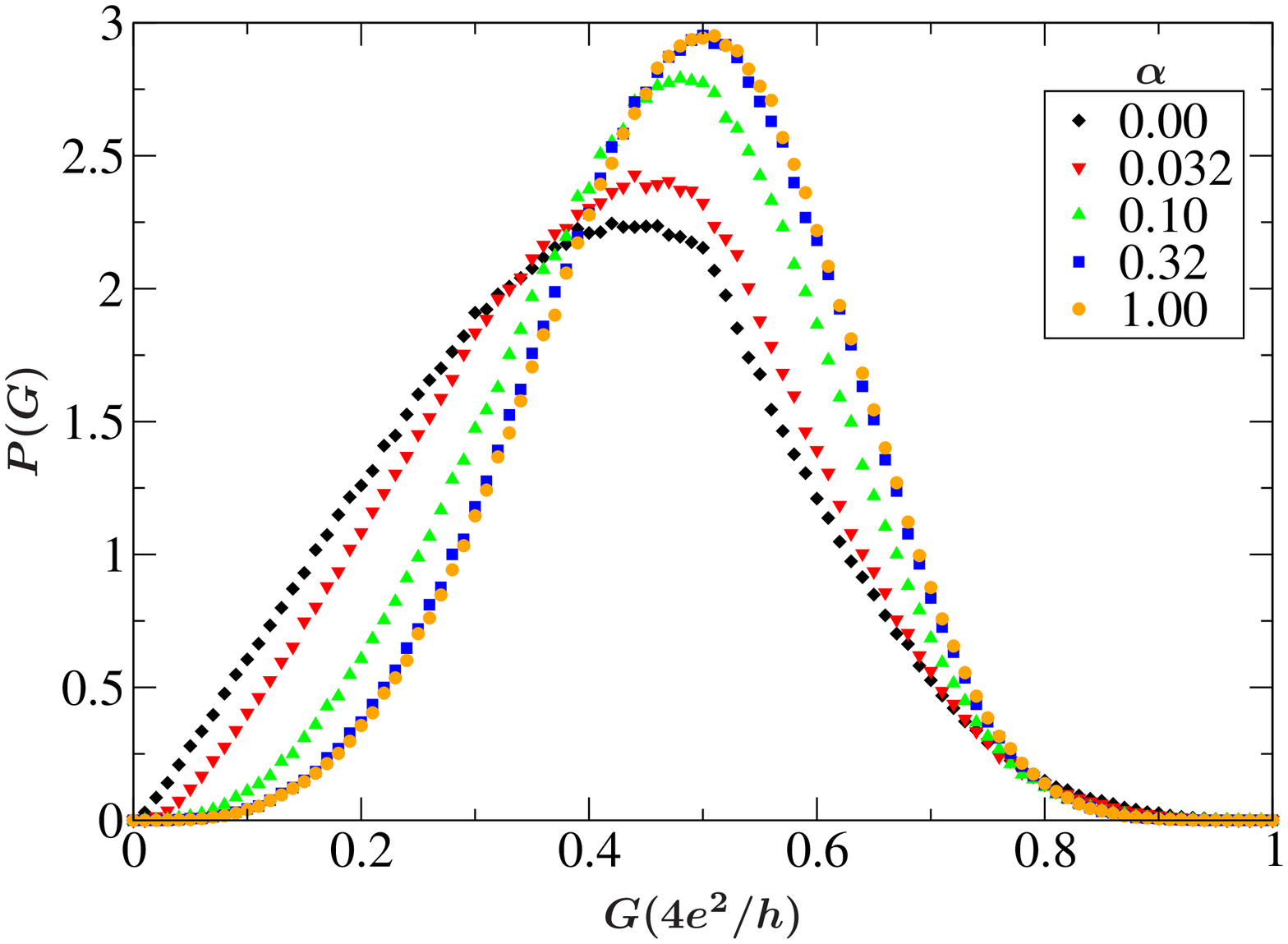}
\includegraphics[width=0.49 \textwidth]{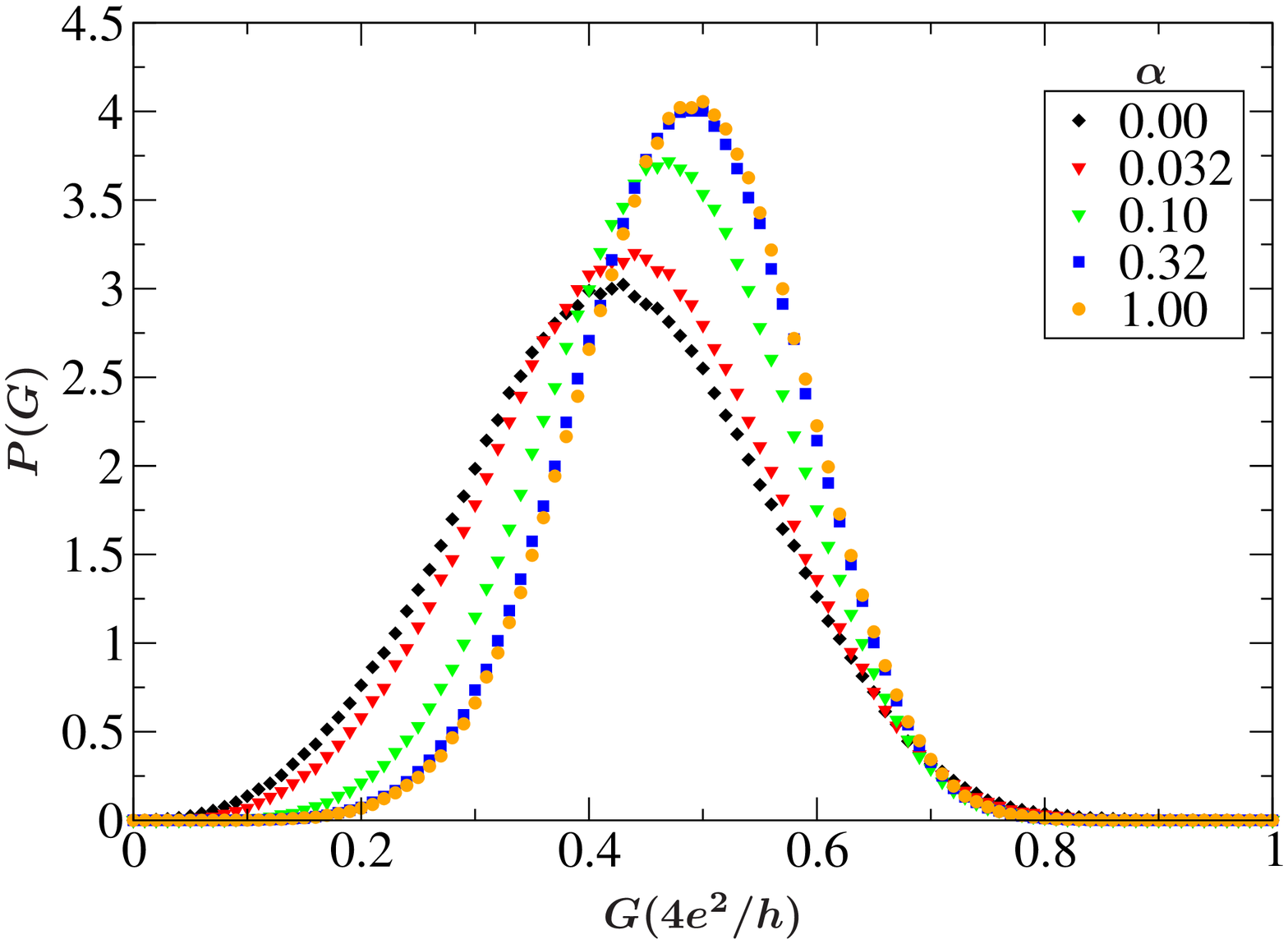}
\caption{Probability distribution of conductances for chaotic Dirac billiards in the extreme quantum regime, $N_1=N_2=1$, with the phase coherence breaking parameters $N_\phi=1$ (left) and $N_\phi=2$ (right). For $\alpha =0,1$ the probability distributions correspond to pure ensembles chGOE and chGUE, respectively.}\label{figura2}
\end{figure*}

After introducing the standard scattering model for the chaotic Dirac billiards connected to three terminals, we may study the time-reversal symmetry and electronic phase coherence breaking. For this purpose, we employ the formulation proposed by B{\"u}ttiker\cite{Buttiker}, which was successful in describing the chaotic quantum billiard as shown in Refs.\cite{Marcus,Mello,Brouwer}. The method was originally used in the framework of random matrix theory, and more recently in Ref.\cite{Petitjean}.

The formulation assumes a third fictitious terminal connected to the chaotic Dirac billiard, which induces the phase coherence breaking or decoherence. The first consistency condition is to maintain null values for the average current in the third terminal, $\left<I_3 \right>=0$, in such a manner that there is only an effective current between the terminals labeled $1$ and $2$. Another consistence condition is the electronic current conservation, $I=I_1=-I_2$. The Landauer-B\"uttiker conductance is changed by those conditions and may be written as \cite{Buttiker,Mello,Brouwer}
\begin{eqnarray}
 G =\frac{2e^2}{h} \left[g_{12}+\frac{g_{31}g_{23}}{g_{31}+g_{32}}\right],\label{g3}
\end{eqnarray}
where $g_{ij}=\textbf{Tr} \left(t_{ij}t_{ij}^{\dagger}\right) $. The open channels in the third lead ($N_\phi=N_3$) are inversely proportional to the dephasing time $\tau_\phi$, while the total open channels $N_1+N_2$ are inversely proportional to the dwell time $\tau_D$ \cite{Barros,Petitjean}. The described open channels satisfy $N_\phi/(N_1+N_2)=\tau_D/\tau_\phi$. The particular limit $N_\phi\gg N_1+N_2$ ( $\tau_D\gg \tau_\phi$) indicates the configuration of electrons remaining longer in the billiard, leading to electronic phase coherence breaking. However, if $N_1+N_2\gg N_\phi$ or, equivalently, $\tau_\phi\gg \tau_D$, the electrons leave the billiard just before losing the coherence.

If the number of open channels in the third lead is null in Eq. (\ref{g3}), we recover the usual Landauer-B\"uttiker conductance,
\begin{eqnarray}
 G =\frac{2e^2}{h} \textbf{Tr} \left(t_{12}t_{12}^{\dagger}\right). \label{landauer}
\end{eqnarray}

\section{Random Matrix Theory Applied to Chaotic Dirac Billiards}
\label{}

\begin{figure*}[!]
\centering
 \includegraphics[width=0.49\textwidth]{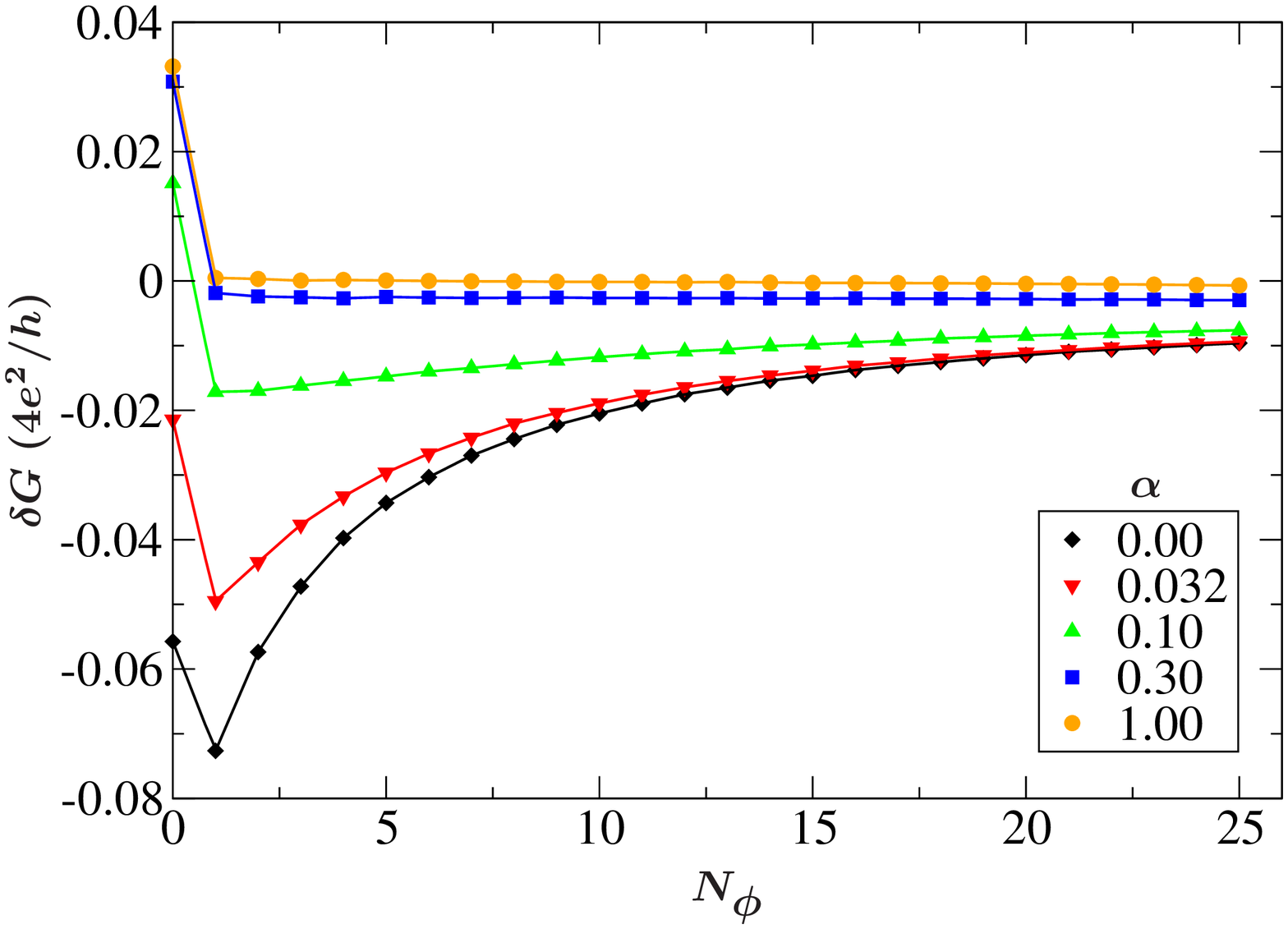}
\includegraphics[width=0.49 \textwidth]{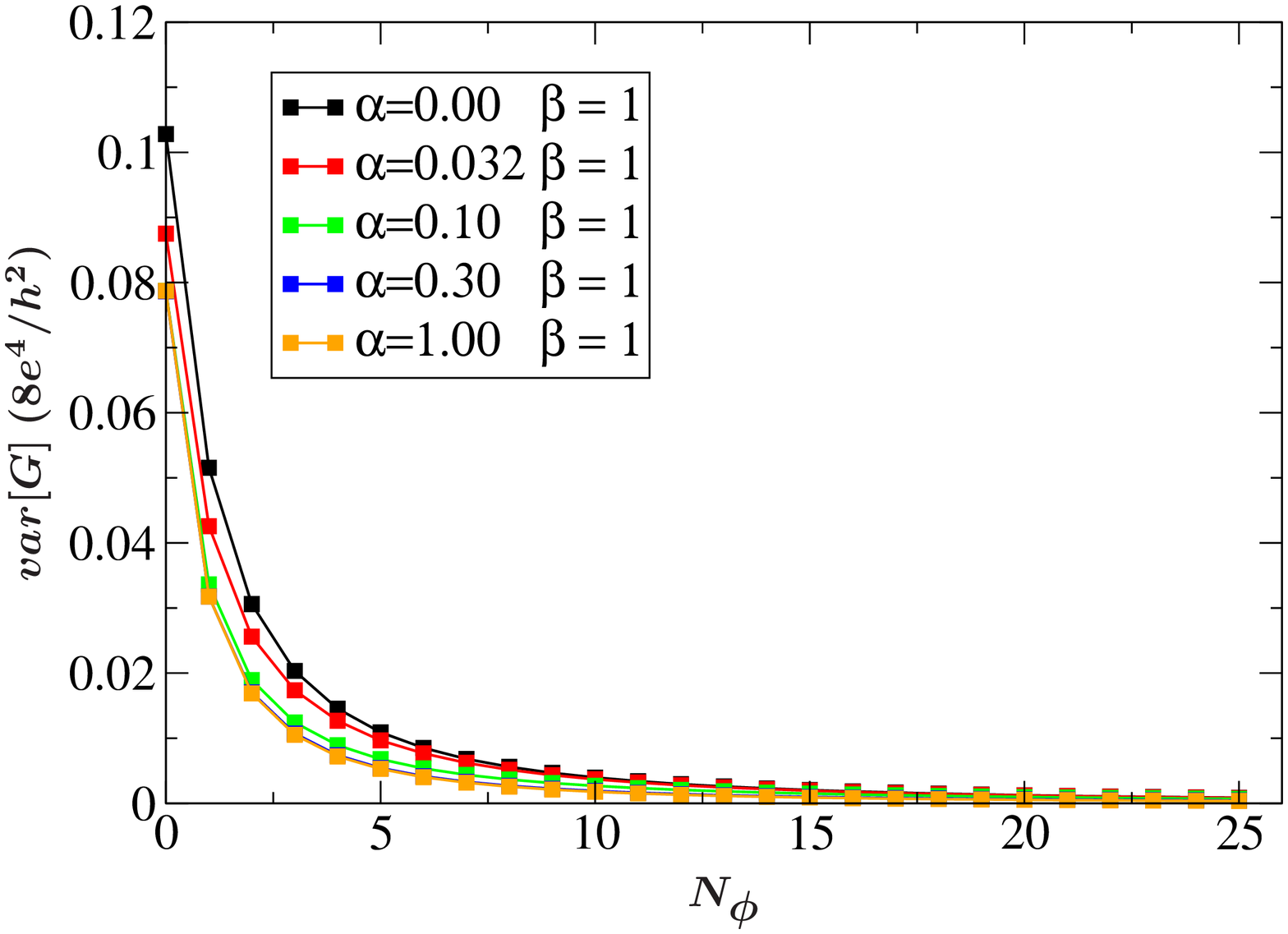}
\caption{Weak localization (left) and variance of conductance (right) of the chaotic Dirac billiard in the extreme quantum regime, $N_1=N_2=1$, as a function of the phase coherence breaking parameter $N_\phi$. For $\alpha =0,1$, they renders the corresponding values for pure ensembles chGOE and chGUE, respectively.}\label{figura3}
\end{figure*}

We start from the anti-commutation relation, Eq.(\ref{H}), which allows to represent the Hamiltonian of the chaotic Dirac billiard in an anti-diagonal form
\begin{equation}
 \mathcal{H}= \left(
 \begin{array}{cc}
  {0} & {\mathcal{T}} \\
 {\mathcal{T}}^{\dagger} & {0}
 \end{array}
 \right).
 \label{ch1}
\end{equation}
The $\mathcal{T}$-block is a random matrix, $M\times M$, which can be decomposed in the form $\mathcal{T} = \mathcal{T}^{(0)} + i\mathcal{T}^{(1)}$. The random matrix theory establishes that the entries of $\mathcal{T}$ can be chosen as members of a Gaussian distribution,
\begin{equation}
P(\mathcal{T})\propto \mathrm{exp}\left[ -\dfrac{M}{2\lambda^2}Tr(\mathcal {TT^\dagger})\right].
\label{ch3}
\end{equation}
The parameter $\lambda =2M \Delta/\pi$ is the variance, related to the electronic single-particle level spacing $\Delta$.

The Eq.(\ref{ch3}) is applicable if the time-reversal symmetry is valid. In this case, the ensemble class is called chiral Gaussian orthogonal ensemble (chGOE). However, if the time-reversal symmetry is broken by a sufficiently intense external magnetic field, the ensemble class is called chiral Gaussian unitary ensemble (chGOE). To analyze the crossover (intermediate values of perpendicular magnetic field) between chGOE and chGUE, we adapt the method proposed in Ref.\cite{Souza} in the context of crossover between the GOE and the GUE (Wigner-Dyson Ensembles).

The matrix $\mathcal{T}$ is a member of the Gaussian ensemble, whose means and variances of their entries can be written as
\begin{eqnarray}
\left\langle (\mathcal{T}_{\mu\nu}^{(0,1)})\right\rangle&=&0, \nonumber\\
\left\langle (\mathcal{T}_{\mu\nu}^{(0)})^2\right\rangle&=&
\dfrac{\lambda^2(1+e^{-2\tau})}{2M},\nonumber\\
\left\langle (\mathcal{T}_{\mu\nu}^{(1)})^2\right\rangle&=&
\dfrac{\lambda^2(1-e^{-2\tau})}{2M}, \, \mu\neq\nu. \label{T}
\end{eqnarray}
The parameter $\tau$ is introduced to break the time-reversal symmetry and, consequently, to control the crossover between chGOE and chGUE. Notice the limits $\tau=0$ and $\tau \rightarrow \infty$ leave the QD to the pure ensembles chGOE and chGUE, respectively.

\section{Numerical Results: Probability distributions of Conductance}
\label{}

\begin{figure*}[!]
\centering
\includegraphics[width=0.6\textwidth]{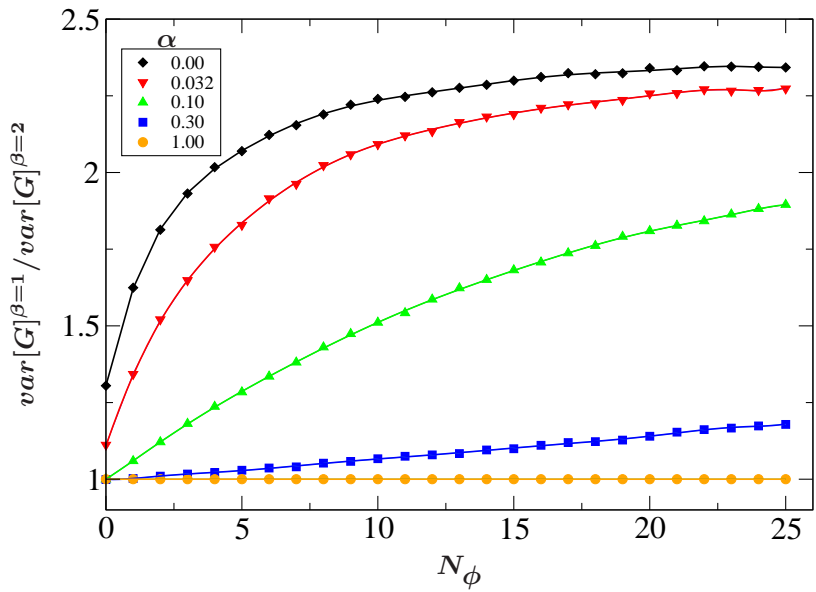}
\caption{The ratio of variance of conductances with (chGOE, $\beta=1$) and without (chGUE, $\beta=2$) time-reversal symmetric for chaotic Dirac billiard in the extreme quantum regime, $N_1=N_2=1$, as a function of the phase coherence breaking parameter $N_\phi$. For $\alpha =0$,  the ratio goes to $2.5$, while for $\alpha =1$ the the ratio is $1$.}\label{figura4}
\end{figure*}

In this section, using Eqs.(\ref{MW}), (\ref{g3}), (\ref{ch1}),  and  (\ref{T}), we numerically obtain the probability distributions of conductance in both the crossover (chGOE $\rightarrow$  chGUE) and in the electronic phase coherence breaking (Quantum $\rightarrow$  Classic) regimes. They are presented as a function of three relevante experimental parameters: (1) the number of open channels in the leads ($N_1$ and $N_2$), (2) the time-reversal symmetry breaking parameter ($\tau$), and (3) the phase coherence breaking parameter ($N_\phi$).

The numerical simulations is implemented using an ensemble of random $ \mathcal{T} $-matrices with order $100 $ ($ M = 100$), and, consequently, with random $ \mathcal{H} $-matrices with order $200 \times 200$ ($200$ resonances), as may be seen in Eq.(\ref{ch1}). To ensure the full convergence of the probability distribution of conductance, we appropriately use $10^6$ realizations of the random $\mathcal{S}$-matrices, Eq.(\ref{MW}). The $\mathcal{W} = \textbf{(}\mathcal{W}_{1}, \mathcal{W}_{2}, \mathcal{W}_{3}\textbf{)} $ matrix has dimension $200\times 2N_{T}$ and describes the coupling of the resonance states of the chaotic Dirac billiard with the propagating modes in the three terminals. This deterministic matrix satisfies a non-direct process, i.e., the orthogonality condition $ \mathcal{W}_{p}^{\dagger}\mathcal{W}_{q}=\frac{1}{\pi}\delta_{p,q}$ holds.
Following Ref.\cite{Souza}, we introduced the parameter $\alpha=\sqrt{\tanh(\tau)}$ assuring numerical convenience. The range of $\tau$ requires $\alpha$ tuned in the interval $0\leqslant \alpha \leqslant 1$.

We begin with a configuration wherein the number of channels is minimal, $N_1=N_2=1$ and $N_\phi =0$. This choice defines the so-called extreme quantum regime and generates a probability distribution of conductance with extreme deviation from the Gaussian. The Fig.(\ref{figura1}), left panel, depicts the probability distribution of conductance for several values of $\alpha$. For $\alpha=0$ and $1$ the probability distributions correspond to chGOE and chGUE, respectively, which are in concordance with the numerical results of Refs.\cite{Barros,Macedo}. However, for intermediate values of $\alpha$, the Fig.(\ref{figura1}), left panel, shows the crossover between pure ensembles, from chGOE to chGUE. The conductance probability distribution on pure ensembles are dramatically affected by small variations of the parameter $\alpha=\{0.032, 0.1\}$, leading to the numerical conclusion that the crossover was completely achieved for $\alpha = 0.30$. Notice the Fig.(\ref{figura1}), left panel, has similar behavior to the Fig.(7) of Ref.\cite{Markos}, obtained by the tight-binding Hamiltonian model of the disordered two-dimensional electron systems with Chiral symmetry. We have shown in the Fig.(\ref{figura1}) that extreme quantum regime on Dirac Billiards have a behavior similar to intermediate regimes ($N_1=2=N_2$) of the Schr\"odinger billiards of Ref.\cite{Souza}. The Fig.(\ref{figura1}), right panel, is plotted in the regime $N_1=N_2=2$ and $N_\phi =0$ and shows the remarkable conclusion that the probability distribution of conductance is only persuasively non-Gaussian in the extreme quantum regime. The Fig.(\ref{figura1}) (right) shows also that Dirac Billiards behavior quickly converges to the Gaussian regime, as compared with Schr\"odinger ones of Ref.\cite{Souza}.

In the Fig.(\ref{figura2}), left panel, we maintain the system in the extreme quantum regime. However, the phase coherence breaking parameter was changed to $N_\phi=1$. We clearly observe the effect of the addition of a single dephasing channel: unlike the Fig.(\ref{figura1}), left panel, the chGUEs probability distribution has a Gaussian (semi-classical) behavior while chGOE remains non-Gaussian. Consequently, the parameter $\alpha$ has a fundamental importance on the full time-reversal symmetry breaking mechanism appearing on the crossover from non-Gaussian (with phase coherence) to Gaussian (without phase coherence) probability distributions of conductance. The Fig.(\ref{figura2}), right panel, was plotted in the extreme quantum regime and $N_\phi =2$, which completely breaks the phase coherence, and generates a full semiclassical (Gaussian) transition to all probability distributions of conductance.

\section{Weak Localization and Universal Fluctuations}

As is well established, in addition to the Ohm semi-classical term, the conductance has quantum interference terms due to the time-reversal symmetry (orthogonal ensemble). These interference terms are generally known as weak localization. In the total absence of time reversal symmetry (unitary ensemble) generated, for example, due to a perpendicular magnetic field, the weak location term disappears. In this section, we investigate the effect of magnetic field on the suppression of the weak localization terms of the conductance. As our statistical simulation indicates, the parameter $\alpha$ naturally plays the role of the magnetic field. Once proven the relevant role of the dephasing in performing the transition from the quantum to the semiclassical regime, we study in detail the consequences of the competition between the parameter $\alpha$ and the dephasing parameter on the interference effects. It is known that the weak localization and the amplitude of the universal fluctuations for the chaotic Dirac billiard reach to zero if $N_\phi\gg N_1+N_2$. We define $\delta G=\left\langle G \right\rangle_{chGOE}-\left\langle G \right\rangle_{chGUE}$ and $var[ G]=\left\langle G^2 \right\rangle-\left\langle G \right\rangle^2$ and fix $\alpha=0$ in the numerical simulation and we confirm this assertion, as depicted in the Fig.(\ref{figura3}). However, for intermediate values of $\alpha$, the weak localization and the amplitude of the universal fluctuations reach zero faster as a function of $N_\phi$. The critical value $\alpha=0.30$ completely breaks both the time-reversal symmetry and the interference effects, regardless of the value of $N_{\phi}$ as demonstrated in some cases in Ref.\cite{Barros}.

In the Fig.(\ref{figura3}), left panel, we show a peculiar transition, a decrease followed by a monotonically increase of the weak localization, in the region of extreme quantum regime, as a function of $N_\phi$. The transition is suppressed as we increment $\alpha$. For $\alpha=0.30$, the weak localization only decreases, from finite values to zero, which is in turn expected for the chGUE. Notice also that the lowest points of the transition are all corresponding to $N_{\phi} = 1 = N_1 = N_2$. For any value of $N_i>1$ this transition disappear, as demonstrated in Ref.\cite{Barros}. This is a very peculiar characteristic of the extreme quantum limit, for which the dephasing performs a preponderant hole.

The Fig.(\ref{figura3}), right panel, shows the monotonically decrease of variance of conductance (in the extreme quantum regime) as a function of $N_\phi$. We do not observe significant differences on the variance in the validity range of $\alpha$. However, we can better understand the variance of conductance by investigating the ratios \cite{Mello} between variance of conductances of chGOE and chGUE, i.e $var[G]_{chGOE}/var[G]_{chGUE}$. This ratio is depicted in Fig.(\ref{figura4}). For $\alpha=0$ the ratio goes to $2.5$ if $N_\phi\gg 1$, as analytically obtained in Ref.\cite{Barros}. However, if we increment the parameter $\alpha$ the ratio goes to values lower than $2.5$ until saturation if $\alpha = 1.0$, yielding exactly the ratio $1$, as expected. Notice also that the previously mentioned critical number, $\alpha=0.30$, renders a ratio different from $1$. That is interesting and a characteristic of the ratio, as there is no significant difference between the probability distribution and variance of conductance for $\alpha=0.30$ and $\alpha=1.0$ (pure ensemble, chGUE), as the Figs.(\ref{figura1}) and Fig.(\ref{figura3}), right panel, indicate.

\section{Conclusion}

In this paper, we perform a complete statistical study of Dirac billiards in several regimes. In particular, we numerically studied the major mechanisms of dephasing and time reversal symmetry breaking. In the transition from the quantum to the semiclassical regime (Gaussian), we glimpse the strong dependence of a myriad of both quantum and universal fingerprints as a function of the number of dephasing channels.

The numerical statistical simulations include a complete study of the universal conductance on Dirac Billiards: the probability distributions, the amplitude of the universal fluctuations and the weak localization term. For the probability distributions of conductance, we found that finite magnetic fields break the time reversal symmetry quickly. We notice that the dephasing plays a dramatic role in the crossover to the semiclassical regime, even for the smallest possible number of open channels (extreme quantum regime). We also study and show the preponderant role of the competition of dephasing and time-reversal symmetry in a myriad of configurations. Our results can be applied to various scenarios involving dephasing fields, ranging from capacitive environments to finite barriers introduced in Dirac billiards. They may also be applied to general systems subjected to tunable fields that perform a crossover between universal ensembles, including chaotic graphene flakes and topological insulators.

\section*{Acknowledgements}
 This work was partially supported by CNPq, CAPES, and FACEPE (Brazilian Agencies).






\begin{thebibliography}{00}


\bibitem{Berry}
M. V. Berry, {\it Some quantum-to-classical asymptotics}, in {\it Chaos and Quantum Physics: Les Houches Lecture} Series 52 (ed. M.-J. Giannoni, A. Voros, and J. Zinn-Justin,North-Holland, Amsterdam, 1991), p. 251.

\bibitem{Bohigas}
O. Bohigas, {\it  Random matrix theories and chaotic dynamics}, in {\it Chaos and Quantum Physics: Les Houches Lecture} Series 52 (ed. M.-J. Giannoni, A. Voros, and J. Zinn-Justin,North-Holland, Amsterdam, 1991), p. 89.

\bibitem{Stockmann}
H.J. St\"{o}ckmann, {\it Quantum Chaos: An Introduction} (Cambridge University Press, Cambridge, 2000).

\bibitem{ARichter}
G. E. Mitchell, A. Richter, and H. A. Weidenm\"uller, Rev. Mod. Phys. {\bf 82}, 2845 (2010).

\bibitem{MelloKumo}
P. A. Mello, and N. Kumar, {\it Quantum Transport in Mesoscopic Systems
Complexity and Statistical Fluctuations},  (Academic, Oxford, 2004).

\bibitem{beenakker}
C. W. J. Beenakker, Rev. Mod. Phys. {\bf 69}, 731 (1997); Rev. Mod. Phys. {\bf 80}, 1337 (2008).

\bibitem{Dietz}
B. Dietz and A. Richter, Chaos {\bf 25}, 097601 (2015).

\bibitem{BeenakkerMarcus}
C. W. J. Beenakker, M. Kindermann, C. M. Marcus, and A.Yacoby, in
Fundamental Problems of Mesoscopic Physics, NATO Science Series II Vol. 154, edited by I. V. Lerner, B. L.
Altshuler, and Y. Gefen (Kluwer, Dordrecht, 2004).

\bibitem{Zirnbauer}
A. Altland and M. R. Zirnbauer, Phys. Rev. B {\bf 55}, 1142 (1997).

\bibitem{JacquodButtiker}
P. Jacquod, R. S. Whitney, J. Meair, and M. Buttiker,  Phys. Rev. B {\bf 86}, 155118 (2012).

\bibitem{mehta}
M. L. Mehta, {\it Random Matrices} (Academic, New York, 1991).

\bibitem{Verbaarschot}
E. V. Shuryak and J. J. M. Verbaarschot, Nucl. Phys. A {\bf 560}, 306 (1993);J. Verbaarschot, Phys. Rev. Lett. {\bf 72}, 2531 (1994).

\bibitem{Mondragon}
M. V. Berry and R. J. Mondragon, Proc. R. Soc. London A {\bf 412}, 53 (1987)

\bibitem{Geim}
L. A. Ponomarenko, F. Schedin, M. I. Katsnelson, R. Yang, E. W. Hill, K. S. Novoselov, A. K. Geim, Science {\bf 320}, 356 (2008).

\bibitem{Mortessagne}
U. Kuhl, S. Barkhofen, T. Tudorovskiy, H.-J.  St\"ockmann, T. Hossain, L. de Forges de Parny, and F. Mortessagne, Phys. Rev. B {\bf 82}, 094308 (2010).

\bibitem{Baranger}
J. Wurm, A. Rycerz, I. Adagideli, M. Wimmer, K. Richter, and H. U. Baranger, Phys. Rev. Lett. {\bf 102}, 056806 (2009).

\bibitem{Grafeno}
J. G. G. S. Ramos, M. S. Hussein, and A. L. R. Barbosa, Phys. Rev. B {\bf 93}, 125136 (2016).

\bibitem{Dietz4}
B. Dietz, A. Richter, and R. Samajda, Phys. Rev. E {\bf 92}, 022904 (2015).

\bibitem{Dietz5}
B. Dietz, T. Klaus, M. Miski-Oglu, A. Richter, M. Wunderle, and C. Bouazza, Phys. Rev. Lett. {\bf 116}, 023901 (2016).

\bibitem{LeiYing}
Lei Ying and Ying-Cheng Lai, Phys. Rev. B {\bf 93} 085408 (2016).

\bibitem{Barros}
M. S. M. Barros, A. J. Nascimento Júnior, A. F. Macedo-Junior, J. G. G. S. Ramos, and A. L. R. Barbosa , Phys. Rev. B {\bf 88}, 245133 (2013)

\bibitem{Klaus}
B. Dietz, T. Klaus, M. Miski-Oglu, and A. Richter, Phys. Rev. B {\bf 91}, 035411 (2015).


\bibitem{beenakkerAndreev}
C. W. J. Beenakker, Lect. Notes Phys. {\bf 667}, 131 (2005).

\bibitem{Mello1}
H. U. Baranger, P. A. Mello , Phys. Rev. Lett. {\bf 73},142 (1994).

\bibitem{Lenz}
G. Lenz, and F. Haake, Phys. Rev. Lett. {\bf 65}, 2325 (1990).

\bibitem{Brouwer1}
A. G. Huibers, S. R. Patel, C. M. Marcus, P. W. Brouwer, C. I. Duru\"oz, and J. S. Harris, Jr. Phys. Rev. Lett. {\bf 81}, 1917 (1998).

\bibitem{Kanzieper}
P. Vidal, E. Kanzieper, Phys. Rev. Lett. 108, 206806 (2012); A. Jarosz, P. Vidal, and E. Kanzieper, Phys. Rev. B {\bf 91}, 180203(R) (2015).

\bibitem{Souza}
A. M. C. Souza, and A. M. S. Mac\^edo, Physica A {\bf 344}, 677 (2004).

\bibitem{Ramos}
J. G. G. S. Ramos, D. Bazeia, M. S. Hussein, and C. H. Lewenkopf, Phys. Rev. Lett. {\bf 107}, 176807 (2011).

\bibitem{nos}
A. L. R. Barbosa, M. S. Hussein, and J. G. G. S. Ramos, Phys. Rev. E {\bf 88}, 010901(R), (2013).

\bibitem{Dietzprb}
B. Dietz, A. Richter, and R. Samajdar, Phys. Rev. B {\bf 92}, 0229004 (2015).

\bibitem{paradox}
A. L. R. Barbosa, D. Bazeia, and J. G. G. S. Ramos, Phys. Rev. E {\bf 90}, 042915 (2014).

\bibitem{nos1}
A. L. R. Barbosa, J. G. G. S. Ramos, A. M. S. Mac\^edo, J. Phys. A: Math. Theor. {\bf 43} 075101 (2010); J. G. G. S. Ramos, A. L. R. Barbosa, A. M. S. Mac\^edo, Phys. Rev. B {\bf 78}, 235305 (2008).

\bibitem{Novaes}
M. Novaes, Annals of Physics {\bf 361}, 51 (2015)

\bibitem{Gopar}
V. A. Gopar and D. Frustaglia, Phys. Rev. B {\bf 77}, 153403 (2008).

\bibitem{Almeida}
F. A. G. Almeida and A. M. C. Souza, Phys. Rev. B {\bf 82}, 115422 (2010).

\bibitem{Berry2}
M.V. Berry and M. Robnik, J. Phys. A {\bf 19}, 649 (1986).

\bibitem{Schanze}
H. Schanze, E. R. P. Alves, C. H. Lewenkopf, and H.-J. St\"ockmann, Phys. Rev. E {\bf 64}, 065201 (2001).

\bibitem{Dietz3}
B. Dietz, T. Friedrich, H. L. Harney, M. Miski-Oglu, A. Richter, F. Schäfer, J. Verbaarschot, and H. A. Weidenm\"uller, Phys. Rev. Lett. {\bf 103}, 064101 (2009).

\bibitem{Buttiker}
M. Buttiker, Phys. Rev. Lett. {\bf 57}, 1761 (1986); Phys. Rev. B {\bf 33}, 3020 (1986).

\bibitem{Marcus}
C. M. Marcus, R. M. Westervelt, P. F. Hopkins, and A. C. Gossard,  Phys. Rev. B {\bf 48}, 2460 (1993).

\bibitem{Mello}
H. U. Baranger, P. A. Mello, Phys. Rev. B {\bf 51}, 4703 (1995).

\bibitem{Brouwer}
P. W. Brouwer, C. W. J. Beenakker, Phys. Rev. B {\bf 51}, 7739 (1995).

\bibitem{Petitjean}
R. S. Whitney, P. Jacquod, C. Petitjean, Phys. Rev. B {\bf 77}, 045315 (2008).

\bibitem{Lewenkopf}
H. Schanze, H.-J. Stöckmann, M. Martínez-Mares, C. H. Lewenkopf, Phy. Rev. E {\bf 71}, 016223 (2005).

\bibitem{Wehling}
T. O. Wehling, A. M. Black-Schaffer, A. V. Balatsky, Adv. Phys. {\bf 76}, 1 (2014).

\bibitem{Silva}
J. G. G. S. Ramos, I. M. L. da Silva, and A. L. R. Barbosa, Phys. Rev. B {\bf 90}, 245107 (2014).

\bibitem{Baranger1}
J. Wurm, M. Wimmer, I. Adagideli, K. Richter, and H. U. Baranger, New J. Phys. {\bf 11} 095022 (2009)

\bibitem{Adagideli}
J. Wurm, K. Richter, and I. Adagideli, Phys. Rev. B {\bf 84}, 075468 (2011); Phys. Rev. B {\bf 84}, 205421 (2011).

\bibitem{Richter}
J. Wurm, M. Wimmer, and K. Richter, Phys. Rev. B {\bf 85}, 245418 (2012).

\bibitem{Richter1}
L. Heße and K. Richter, Phys. Rev. B {\bf 90}, 205424 (2014).

\bibitem{Stampfer}
F. Libisch, C. Stampfer, and J. Burgd\"orfer, Phys. Rev. B {\bf 79}, 115423 (2009).

\bibitem{Grebogi}
R. Yang, L. Huang, Ying-Cheng Lai, and C. Grebogi, EPL {\bf 94}, 40004 (2011).

\bibitem{Rycerz}
A. Rycerz, Phys. Rev. B 85 245424 (2012); Phys. Rev. B {\bf 87} 195431 (2013).

\bibitem{Markos}
P. Markos, L. Schweitzer, Physica B {\bf 407}, 4016, (2012).

\bibitem{Macedo}
A. F. Macedo-Jr., A. M. S. Mac\^edo, Phys. Rev. B {\bf 77}, 165313 (2008); Phys. Rev. B {\bf 66}, 041307 (2002).




\end{thebibliography}



\end{document}